\documentclass[12pt]{iopart}
\usepackage{iopams}  
\usepackage[dvips]{graphicx}
\usepackage{wrapfig}
\begin{document}

\title[ECRH plasma in the LHD]{Extension and its characteristics of ECRH plasma in the LHD}

\author{
S~Kubo,
T~Shimozuma,  
Y~Yoshimura,
T~Notake,
H~Idei\dag,
S~Inagaki,
M~Yokoyama,
K~Ohkubo,
R~Kumazawa,
Y~Nakamura,
K~Saito,
T~Seki,
T~Mutoh,
T~Watari,
K~Narihara,
I~Yamada,
K~Ida,
Y~Takeiri,
H~Funaba,
N~Ohyabu,
K~Kawahata,
O~Kaneko,
H~Yamada,
K~Itoh,
N~Ashikawa,
M~Emoto,
M~Goto,
Y~Hamada,
T~Ido,
K~Ikeda,
M~Isobe,
K~Khlopenkov,
T~Kobuchi,
S~Masuzaki,
T~Minami,
J~Miyazawa,
T~Morisaki,
S~Morita,
S~Murakami\ddag,
S~Muto,
K~Nagaoka,
Y~Nagayama,
H~Nakanishi,
Y~Narushima,
K~Nishimura,
M~Nishiura,
N~Noda,
S~Ohdachi,
Y~Oka,
M~Osakabe,
T~Ozaki,
B~J~Peterson,
A~Sagara,
S~Sakakibara,
R~Sakamoto,
M~Shoji,
S~Sudo,
N~Takeuchi,
N~Tamura,
K~Tanaka,
K~Toi,
T~Tokuzawa,
K~Tsumori,
K~Watanabe,
T~Watanabe,
K~Yamazaki,
M~Yoshinuma,
A~Komori
and O~Motojima
}

\address{\ National Institute for Fusion Science, 
                Toki, Gifu 509-5292, Japan}

\address{\dag\ Advanced Fusion Research Center, Research Institute for Applied Mechanics, 
Kyushu University, Kasuga, 816-8580, Japan}

\address{\ddag\ Department of Nuclear Engineering, 
Kyoto University, Kyoto, 606-8501, Japan}

\begin{abstract}

One of the main objectives of the LHD is to extend the plasma
confinement database for helical systems and to demonstrate such 
extended plasma confinement properties to be sustained in steady state. 
Among the various plasma parameter regimes, the study of confinement properties in the
collisionless regime is of particular importance.
Electron cyclotron resonance heating (ECRH) has been extensively
used for these confinement studies of the LHD plasma from the initial operation.
The system optimizations including the modification of the 
transmission  and antenna system are performed with the special emphasis on
the local heating properties. 
As the result, central electron temperature of more than
10 keV  with the electron density of 0.6 $\times$ 10$^{19}$ m$^{-3}$
is achieved near the magnetic axis. The electron temperature profile is characterized by 
a steep gradient similar to those of an internal transport barrier observed in 
tokamaks and stellarators. 168 GHz ECRH system demonstrated efficient heating at
over the density more than 1.0 $\times$ 10$^{20}$ m$^{-3}$.
CW ECRH system is successfully operated to sustain 756 s discharge.

\end{abstract}



\maketitle
%
%
%
\section{Introduction}
  Electron cyclotron resonance heating (ECRH) plasmas
  have been extensively
  used for the confinement study
  from the initial operation of the LHD,
  where, extending the plasma
  confinement database of helical systems and  
  demonstrating  real steady state plasma confinement
  are its main subjects
  \cite{Iiyoshi,Motojima}. 

  Recently, the ECRH system is upgraded to operate 
  four 168 GHz, two 84 GHz, and  two 82.7 GHz  gyrotrons, simultaneously.
  An 84 GHz CPD diode CW gyrotron 
  with diamond window is also operated to perform long pulse discharge.
  As the result of system optimizations, including those of injection focal points, 
  and polarizations, central electron temperature 
  of more than 10 keV  with the electron density of 0.6 $\times$ 
  10$^{19}$ m$^{-3}$  is achieved near the magnetic axis\cite{Kubo02a}. 
  The electron temperature profile is characterized by 
  a steep gradient similar to those of an internal transport 
  barrier (ITB) observed in 
  tokamaks and stellarators.
  The extension of the plasma  parameter in such low collisional regime 
  accelerated the study on the structure formation 
  of the electron temperature 
  and its relation to the  
  radial electric field expected from neoclassical transport theory 
  and resultant ITB in the LHD\cite{ShimozumaPPCF}.  

Opposite direction to the low density high temperature ECRH is high density heating.
 The effect of heating is confirmed up to the 1.5 $\times$ 10$^{20}$ m$^{-3}$ .
 Second harmonic ECRH at 168 GHz is proven to be effective for such high density, 
 where the fundamental ECRH at 84 GHz is not accessible due to the cutoff.
 
  Continuous or the long time sustainment of the plasma is
  another important missions for the LHD.
  The ECRH system is upgraded to demonstrate the continuous plasma
  sustainment in LHD. This upgrade includes setting and operation of 
  84 GHz CW gyrotron, the enforcement of
  the cooling for the waveguide transmission line, and
  installation of a waveguide antenna system.  Injection of
  about 70 kW demonstrated that the plasma 
  with the time and line averaged density of 2.4 $\times$ 10$^{17}$ m$^{-3}$ could
  be sustained for more than 750 sec.
\section{Electron Cyclotron Heating System in LHD}
\begin{figure}[tbp]
    \begin{center}
      \resizebox{150mm}{!}{\includegraphics{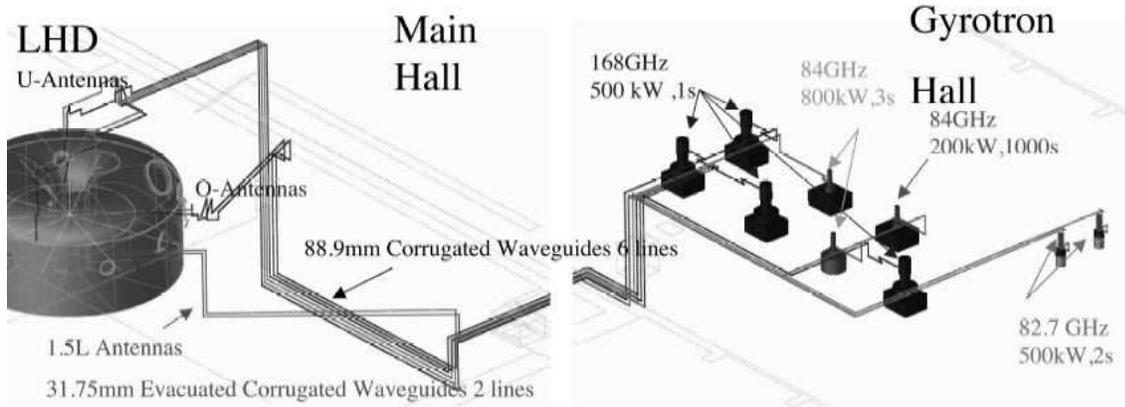}}
      \caption{ Bird eye's view of all ECRH system for LHD. 4-168 GHz, 
2-84 GHz, 2-82.7 GHz gyrotrons are operated simultaneously. 6-88.9 mm, and 
2-31.75 mm corrugated waveguide system transmit the power from gyrotron 
hall to the LHD. Both 31.75 waveguide system are evacuated and one of them 
is used for long pulse experiment. }
      \label{fig:ECHsystem}
    \end{center}
  \end{figure}

Bird eye's view of  ECH system in LHD operated during last 
  experimental campaign is shown in Fig. \ref{fig:ECHsystem}.  
  Six transmission lines of 88.9 mm id  and two of 31.75 mm id corrugated waveguide,
  with total length of about 100 meters for each are in operational
 \cite{Kubo99,Shimozuma01}.
  \begin{table}
    \caption{ List of operated gyrotron and transmission lines in LHD. Here, CPD means potential depressed collector.} 
    \begin{tabular}{lcccccccc}
      \hline
      {\footnotesize Gyrotron No.}
      & {\# 1}
      & {\# 2}
      & {\# 3}
      & {\# 7}
      & {\# 4} 
      & {\# 5}
      & {\# 11}
      & {\# 12}\\
                {\footnotesize  freq. (GHz)}
                & {168}& {168}& {168} & {168}
                & {84} & {84}
                & {82.7} & {82.7}\\
                {\footnotesize manufacturer}
                & {\tiny  Toshiba} & {\tiny  Toshiba} & {\tiny  Toshiba} & {\tiny  Toshiba}
                & {\tiny  Gycom} & {\tiny  Gycom}
                & {\tiny  Gycom} & {\tiny  Gycom}\\
                & {\tiny  Triode}  & {\tiny  Triode} & {\tiny  Triode} & {\tiny  Triode}
                & {\tiny  Diode} & {\tiny  Diode}  
                & {\tiny  Diode} & {\tiny  Diode}\\ 
                {\footnotesize spec. power (kW)}
                & {500} & {500} & {500} & {500}
                & {800} & {800}
                & {500} & {500}\\
                {\footnotesize pulse width (s)}
                & {1} & {1} & {1} & {1}
                & {3} & {3}
                & {2} & {2}\\
                \hline
                    {\footnotesize  Power Supply} 
                    & {\tiny  \#1 CPD }  & {\tiny  \#1 CPD } & {\tiny  \#1 CPD } & {\tiny  \#3 CPD }
                    & {\tiny  \#2 CPD}  & {\tiny  \#2 CPD}  
                    & {\tiny  Non CPD } & {\tiny  Non CPD }\\ 
                    \hline
                        {\footnotesize  waveguide dia.(mm)} 
                        & {88.9} & {88.9} & {88.9} & {88.9} 
                        & {31.75}  & {31.75} 
                        & {88.9} & {88.9}\\
                        & {\tiny dry air} & {\tiny dry air} & {\tiny dry air} & {\tiny dry air} 
                        & {\tiny evacuated} & {\tiny evacuated} 
                        & {\tiny dry air} & {\tiny dry air}\\
                        {\footnotesize total length (m)}
                        & {92} & {92} & {78} & {94}
                        & {65} & {72}
                        & {116} & {115}\\
                        {\footnotesize No. of Bends}  
                        & 18 & 21 & 16 & 21 
                        & 10  & 10 
                        & 19  & 15\\
	                \hline
	                    {\footnotesize Max. P$_{in}$ (kW)}
	                    & 180 & 212 & 186 & 160
                            & 383 & 408
                            & 254 & 286\\	
	                    {\footnotesize Max. width (s)}
	                    & 1.0 & 1.0 & 1.0 & 0.9
                            & 1.5 & 1.5 
                            & 1.5 & 1.5\\	
                            \hline
	                    {\footnotesize Loss at MOU (\%)}
	                    & 23.0 & 24.0 & 14.0 & 32.8
                            & 10.0 & 10.0
                            & 8.5 & 6.5\\	
	                    {\footnotesize Trans. Efficiency (\%)}
	                    & 69.1 & 85.8 & 84.8 & 79.9
                            & 66.7 & 66.7 
                            & 95.6 & 89.3\\	
                            \hline
    \end{tabular}
    \label{tab:TransList}
  \end{table}
  In Table \ref{tab:TransList} are also shown the power loss rate 
  at the MOU and transmission efficiency for each line.
  It should be noted that the transmission 
  efficiency for \#11 and  \#12 are almost 90 \%, although 
  the number of miter bends and total path length are large.
  After this optimization, the threshold power for the arcing 
  inside the waveguide appreciably increased.
  Relatively low efficiency for 168 GHz systems (\#1-3,\#7) 
  and scatter of the values may be due to the sensitivity 
  of the alignment of the input beam axis to that of waveguide.  
  Low transmission efficiency for 31.75 mm diameter evacuated 
  waveguides (\#4 and \#5) 
  may be attributed to the low purity of the coupled HE$_{11}$ mode. 
  Since the axis alignment is less critical in the small diameter
  waveguide system, further optimization of the MOU mirrors or 
  adjustment of the position of the 
  waveguide mouth might be necessary.
  The power from each gyrotron is transmitted 
  through a corrugated waveguide system and injected by a 
  quasi-optical antenna system.  
  Two sets of U (upper ) port antenna consist of
  two sets of mirrors for 82.7 and 168 GHz. 
  The antenna mirrors are designed using the phase constant 
  method assuming Gaussian optics\cite{Kubo95}.  
  Designed beam waist sizes on the
  mid-plane of LHD are 15 and 50 mm in radial and toroidal 
  directions, respectively.
  These values and its steerability on the mid-plane of the 
  LHD are confirmed by low power test.
  In each antenna system, final plasma facing mirror is 
  plane mirror and can be steered 
  by remote-controlled super sonic motors around two axis 
  ( azimuth and elevation angle ).
  
  Due to the complexity of the configurations of the transmission 
  system, quasi-optical antenna system, the optimum setting of the 
  antenna angles and polarizer rotation angles are not 
  straightforward.  Furthermore, the definition of the injection 
  parameters on the frame of the LHD magnetic configuration is 
  necessary and the actual quick selection and setting of these 
  angles are required depending on the various experimental 
  purposes. 
  In order to perform effective local heating by the electron 
  cyclotron waves, it is necessary to select more effective heating 
  mode on the resonance between two eigen modes in the plasma.  
  Desired polarization state to excite the effective mode is determined by the 
  angle between injection and magnetic field directions at the interface 
  of the injected beam to the plasma.  The polarization state at the 
  interface is controlled by the combination of rotation angles 
  of two polarizers and antenna angles.  Given the magnetic field 
  configuration and the magnetic field strength, the optimum 
  antenna setting angles can be determined using a geometrical 
  configuration of the steering antenna.  
  Once the injection angle of the beam is determined, 
  the local magnetic field strength and the angle 
  between injection and magnetic field direction at the 
  interface of the plasma can be calculated.  
  Using such derived magnetic field strength, the angle and the mode, 
  the necessary polarization state can be calculated by the cold plasma 
  dispersion relation.  This polarization state  is 
  projected back to the polarizers to determine the best combination
  of the two polarizers.
%
  \begin{figure}[tbp]
    \begin{minipage}{0.45\textwidth}
      a)
      \begin{center}
        \includegraphics[width=5.0cm,clip]{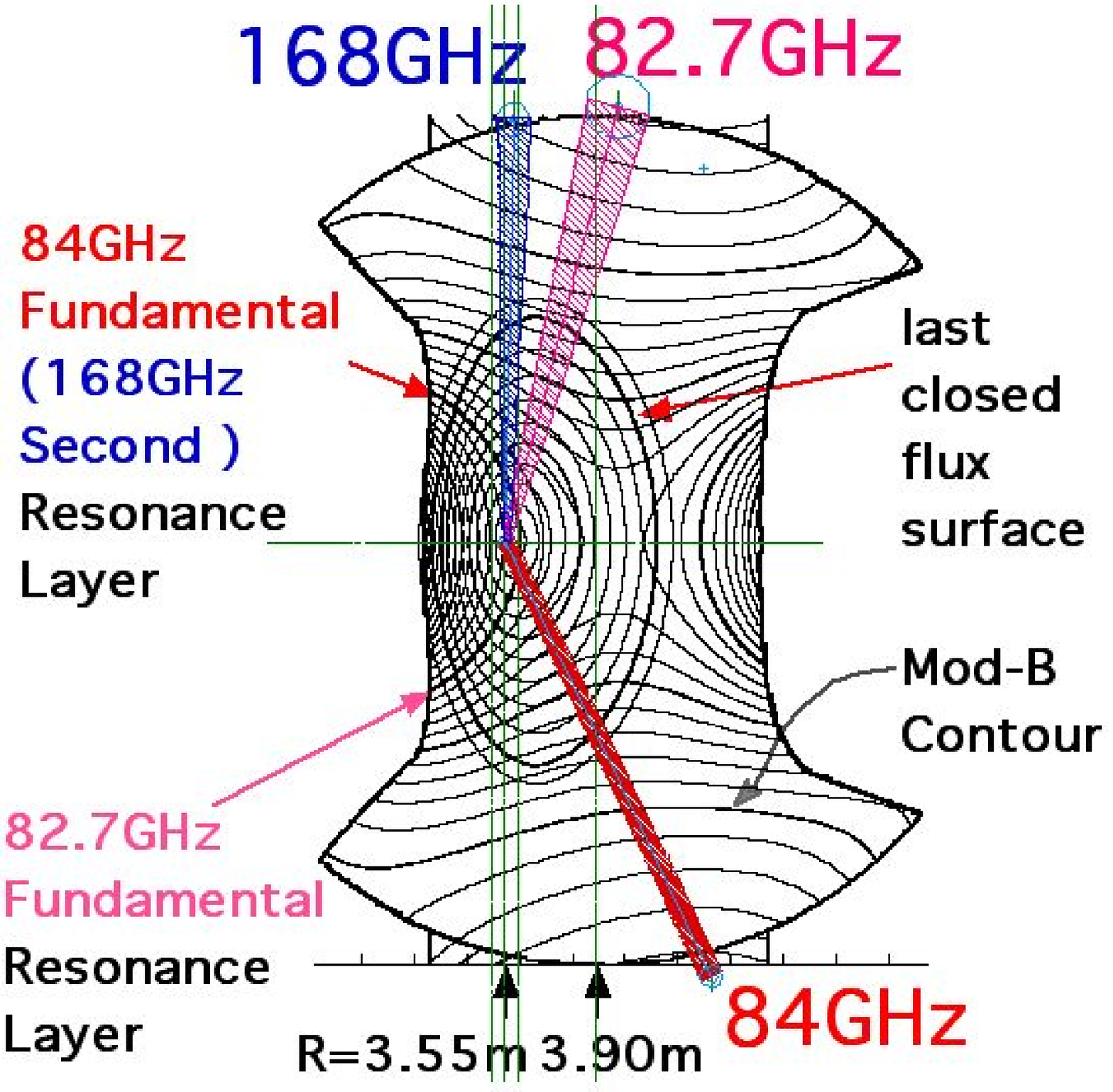}
        \label{fig:AntennaSystem}
      \end{center}
    \end{minipage}
    \begin{minipage}{0.02\textwidth}
    \end{minipage}
    \begin{minipage}{0.48\textwidth}
      b)
      \begin{center}
        \includegraphics[width=5.0cm,clip]{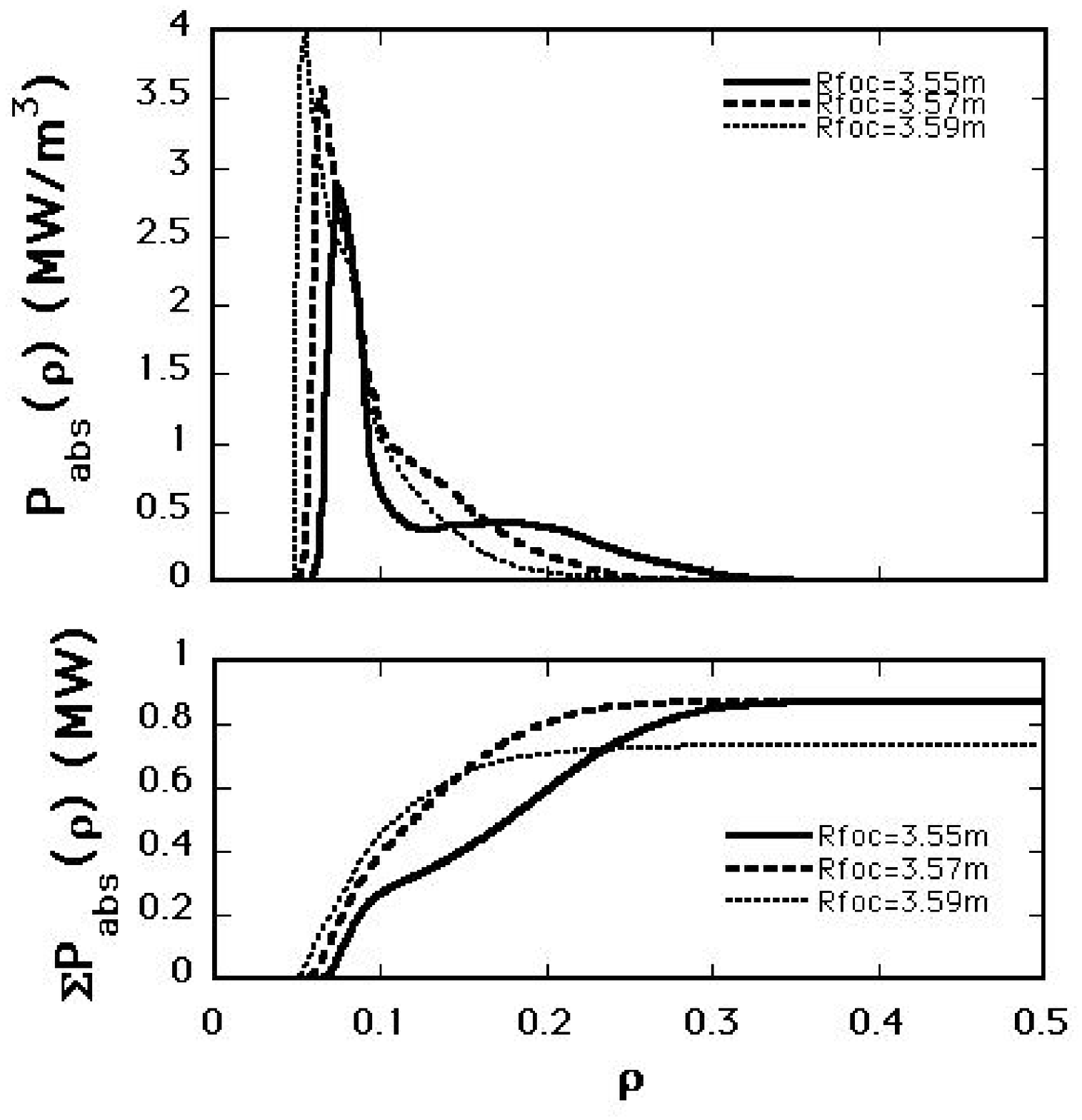}
      \end{center}
    \end{minipage}
        \caption{ a) Flux surfaces and mod-B 
          contours in LHD at vertically elongated
          cross section. Injected microwave beam from 
          upper and lower antennas are
          shown with the beam waist size in scale.  
          b)Expected power deposition 
          profile calculated from
          ray tracing. Power from horizontal antenna 
          is excluded in this calculation.}
        \label{fig:raydep}
  \end{figure}

  Fig. \ref{fig:raydep} a)
  shows the injection beams,
  mod-B contours, and flux surfaces in  
  the vertically elongated poloidal
  cross section. Two sets of upper beams, 
  one lower beam from a vertical
  antenna, and two beams from horizontal antennas are used.    
  In order to attain high electron temperature 
  at center, the magnetic field strength and the confinement magnetic 
  field configuration 
  are selected to have a power
  deposition as nearly on-axis as possible. 
  One of the optimum  combinations is the magnetic axis at 3.53 m and the 
  toroidally averaged magnetic field strength on the axis at 2.951 T. 
  The expected power deposition profile estimated by ray tracing code, including 
  the weakly relativistic effect\cite{Kubo02b,Bornatici83}, indicates 
  that almost all of the injected power from the upper and
  lower antennas are concentrated within an 
  averaged minor radius of $\rho$
  $\approx$ 0.2 as shown in 
  Fig. \ref{fig:raydep} b)
  , here, three cases of 
  different injection angle is plotted to see 
  the allowances and errors in
  antenna setting. 
  Lower traces are integrated absorption power.
  Almost 100\% power absorption can be expected 
  in the density, and temperature
  regime discussed in this paper, provided 
  that the beams crosses the resonance in the plasma
  confinement region and proper injection polarization state
  is selected.
 %
  
  %
  \section{Low collisionality regime}
 The magnetic field at the magnetic axis is effectively increased by shifting 
  the axis inward (R$_{axis}$=3.5 - 3.6) from 
  the standard position( R$_{axis}$= 3.75 m ) to locate 
  the cyclotron layer across the axis.  The injection beam aligned carefully, 
  to hit the resonance on the magnetic axis.
  As a result, ECH beams were concentrated near the shifted axis.
  For the on-axis heating, 
  strongly focussed Gaussian
  beams at the fundamental and second harmonic resonances
  are directed to the resonances near the magnetic axis.
  The microwave sources used in this case were  84 GHz, 
  two 82.7 GHz, and three 168 GHz
  gyrotrons. 
    Fig. \ref{fig:AntennaSystem}
  shows the injection beams,
  Mod-B contours, and flux surfaces in  the vertically elongated poloidal
  cross section. 
  Two sets of upper and one lower beams from a vertical
  antenna, and two beams from horizontal antennas are used.    
  The magnetic field strength and the configuration are selected to have power
  deposition as nearly on-axis as possible. 
  The selected magnetic axis is 3.53 m and the toroidally averaged magnetic 
  field strength on the axis is 2.951 T. The expected power deposition
  profile is shown in Fig. \ref{fig:raydep}.
  \begin{figure}[tbp]
    \begin{minipage}{0.5\textwidth}
      \begin{center}
        \includegraphics[width=80mm,clip]{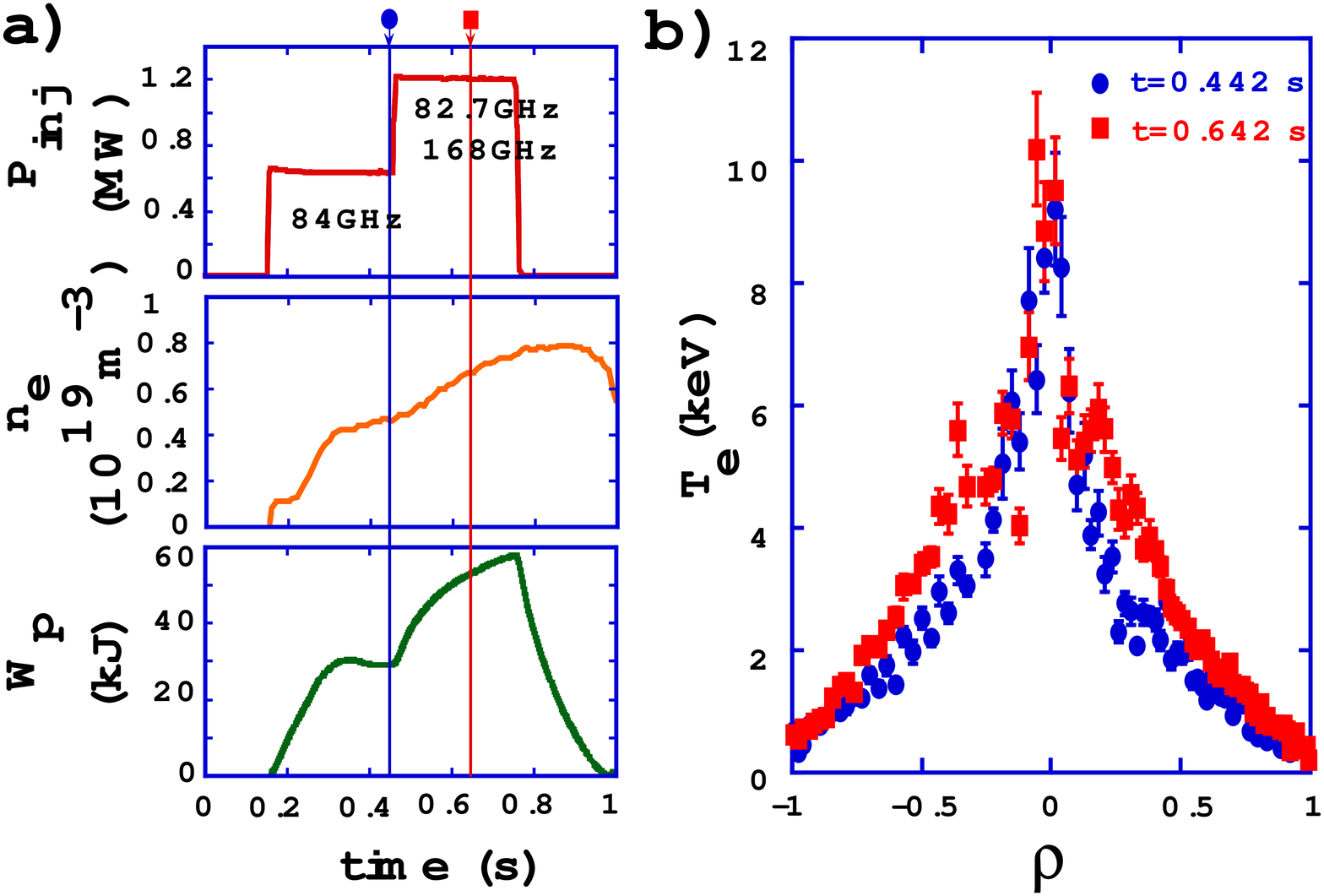}
        \caption{a)Time evolution of injected ECH power (upper), electron
            density (middle) and stored energy (bottom).
            b) T$_{e}$ profile measured at each timing indicated by the arrow in a). }
        \label{fig:highTe}
      \end{center}   
    \end{minipage}
    \begin{minipage}{0.02\textwidth}
    \end{minipage}
    \begin{minipage}{0.48\textwidth}
      \begin{center}
        \includegraphics[width=80mm,clip]{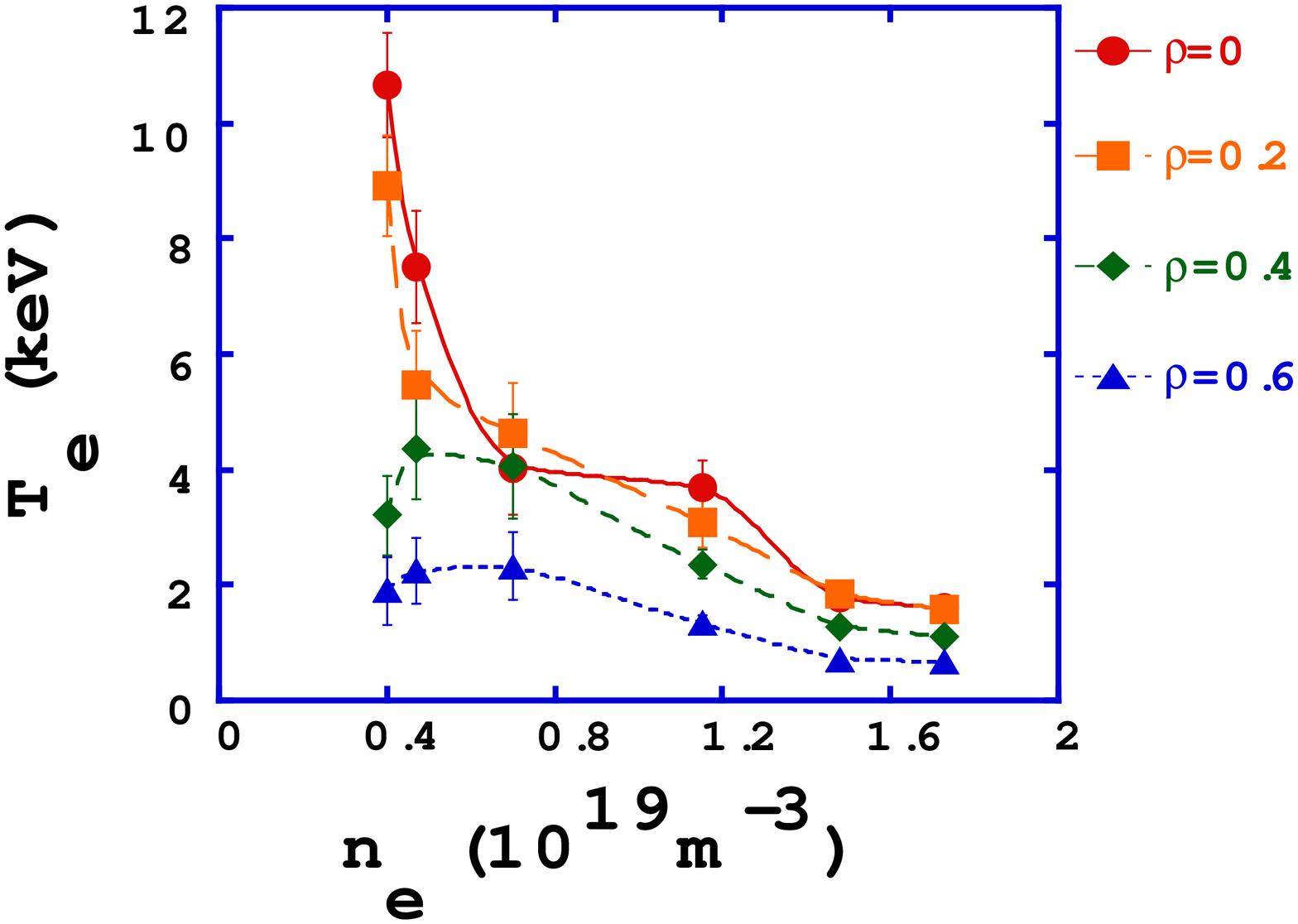}
        \caption{Dependence of T$_{e}$ at $\rho$=0,0.2,0.4 and 0.6 on the electron density.}
        \label{fig:nedep}
      \end{center} 
    \end{minipage}
  \end{figure}
  Fig. \ref{fig:highTe} a)
   shows the evolution of plasma parameters and
  temperature profiles when the central electron T$_{e0}$ exceeded 10 keV. 
  The time evolution of the injected total power, the electron density, 
  and the stored energy are shown here for the shot when the highest central 
  electron temperature is recorded in the LHD. 
  Almost 1.2 MW ECH power is concentrated inside $\rho$ $\approx$ 0.2.
  The electron density is slightly increasing but stays 0.5 to 0.6
  $\times$ 10$^{19}$m$^{-3}$ in this case.
  
  The two 84 GHz gyrotrons injected 0.7 MW to produce and heat
  the plasma.  After the density and the stored energy had attained a quasi-
  steady state, 
  the 168 GHz and 82.7 GHz power are added simultaneously.
  Although no additional gas puff is supplied, the density keeps increasing
  slightly. A high power YAG-Thomson scattering system is used at
  the times indicated by the arrows in 
  Fig. \ref{fig:highTe} 
  a).
  The profile is already sharp in the phase when only the 84 GHz power is
  injected.  The 82.7 GHz power raises the central electron temperature to
  more than 10 keV.  
  These high electron temperature modes appear only when the injected power
  exceeds a certain threshold level, and this threshold level increases with the
  electron density.
  Figure \ref{fig:nedep}
  shows the dependence of the local electron temperature at 
  $\rho$ = 0, 0.2, 0.4 and 0.6 on the averaged electron density under the same
  injection condition. Since the expected power deposition and the
  deposition profiles do not change significantly, the sharp
  increase in the electron temperature at  $\rho$ = 0, 0.2  suggests
  the presence of some non-linear mechanism.
  The feet of the ITB are more prominent when the ECH is applied 
  on counter NBI target plasma.
  It is clear that the threshold power depends on the electron density. 
  Other observations with respect to the ITB are as follows: 
  1) The power deposition profile plays an important role in the formation
  of the ITB. The total deposited power inside $\rho < $ 0.2 seems the
  key factor. 
  2) Reduction of the ion and impurity transport after 
  the ITB  
  formation 
  are suggested from impurity observations.
  indicates the presence of a positive electric field. 
  3) The location of the feet of the ITB depends on the direction of the NBI, 
  which may be related to the position of the rational surface $\iota /
  2\pi$= 0.5\cite{Takeiri}. The $\iota / 2\pi $ =0.5 surface plays an
  important role in expanding this interface outside.  
  The clear ITB or foot point near $\rho $ = 0.2 to 0.3 is observed 
  in the case of counter NBI as a target plasma for ECRH. 
  The slope of the temperature profile gradually increases 
  and clear foot point disappears in the case of co-NBI.  
  4) The density and electron temperature region of electron root 
  predicted from neoclassical theory coincides with the
  region where the ITB appears.

  These facts support the proposed anomalous transport reduction mechanism
  which the  radial electric field shear reduces the fluctuation level
  and the shear itself is formed by the radial interface between electron
  and ion roots \cite{Fujisawa}. 
  The details of the formation of the ITB is under study 
  using heat pulse propagation. Here, the ITB is formed mainly by
  high power fundamental heating (84 and 82.7 GHz ) and the
  modulated second harmonic heating is used as a perturbation source.  
  The drastic difference in the transient transport between 
  with and without ITB is observed.  The time lag of the heat pulse
  propagation clearly changes across the ITB region. 
  This analysis is also enabled by the optimization 
  of the modulating 168 GHz injection parameter. 

  The forward and backward transitions between ITB and normal confinement 
  occurred when a part of the 
  ECRH power is turned off and on to cross the threshold power
  for the ITB\cite{ShimozumaPPCF}.
  The decay time of the central temperature 
  just after crossing down the threshold power is apparently 
  long and high temperature gradient is kept for 60 ms. On the other 
  hand, 
  the decay time is short when the ECRH power kept below threshold.
  When the power increased stepwise crossing the threshold, high temperature
  gradient region grows gradually and the temperature profile 
  jumps to form ITB about 40 ms after the stepwise increase.
  These results suggest that the temperature gradient helps to lower the
  threshold power for ITB formation\cite{yokoyama}.
  
%
\section{High Density ECH}
\begin{figure}[tbp]
    \begin{minipage}{0.5\textwidth}
      \begin{center}
	      \resizebox{70mm}{!}{\includegraphics{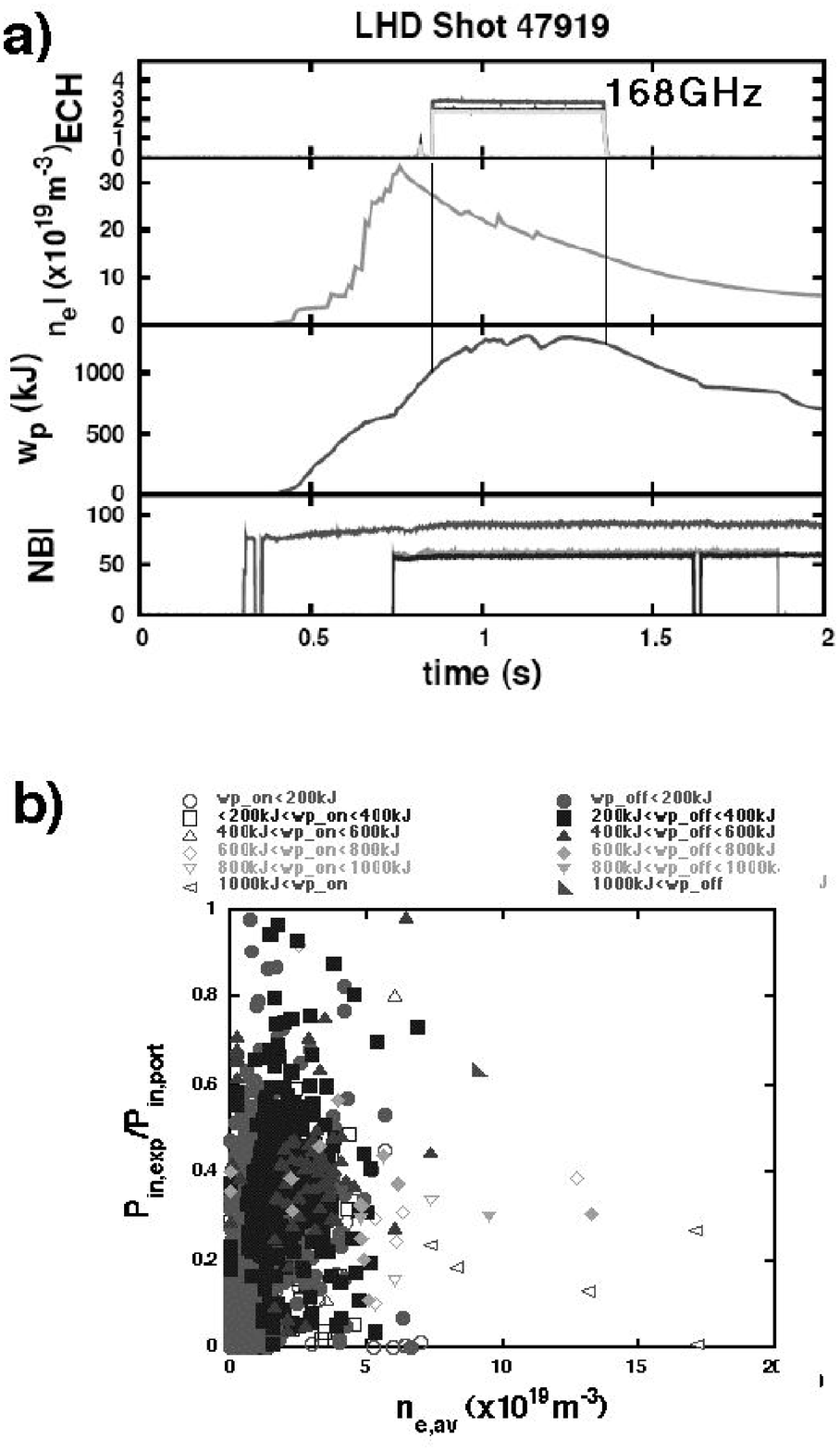}}
      \caption{a)Effect of 168 GHz second harmonic heating at high density plasma where the highest stored energy of 1.3MJ is recorded. b)Dependence of the heating efficiency of 168 GHz injection on target plasma density}
      \label{fig:Efficiency}
	  \end{center}
    \end{minipage}
    \begin{minipage}{0.02\textwidth}
    \end{minipage}
    \begin{minipage}{0.48\textwidth}
      \begin{center}
      \resizebox{60mm}{!}{\includegraphics{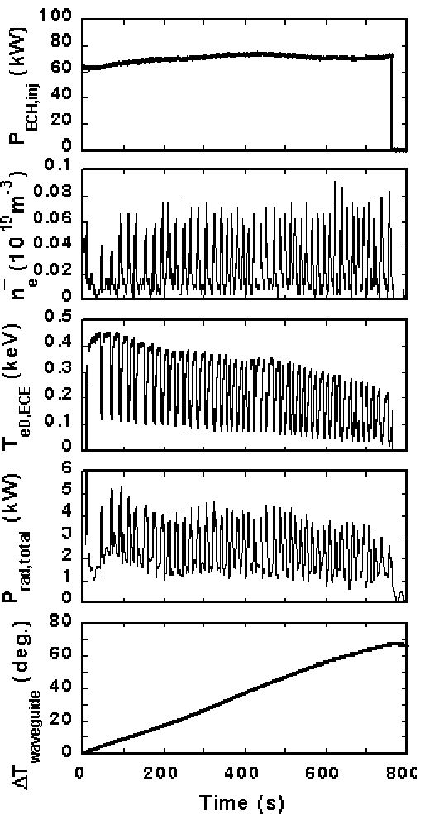}}
      \caption{Time behavior of plasma parameters during 
          756 sec injection.
        a) Injected power,  b) averaged electron density, 
        c) ECE radiation temperature,
        d) Plasma radiation and b) temperature rise of a part 
        of waveguide component, respectively}
      \label{fig:CWplasma}
      \end{center} 
    \end{minipage}
  \end{figure}

Second harmonic high frequency ECH is attractive because of its high density limit.
The second harmonic heating system at 168 GHz for  3 T has been prepared to perform
high density plasma heating. 
Second harmonic heating has its advantage in the accessibility 
in the high density region.  Clear and definite demonstration of the effect is the 
increase of stored energy at the high density above the cut off of 84 GHz fundamental heating.
The cut off density of 84 GHz fundamental O mode is 0.88 $\times $ 10$^{20}$ m$^{-3}$ while that of
168 GHz second harmonic X mode is 1.75 $\times$ 10$^{20}$ m$^{-3}$.
Actually, 168 GHz ECRH power is applied on the high density, high stored energy plasma and
helped to sustain and push highest record value. In Fig. \ref{fig:Efficiency} a) is shown  the 
wave forms of the shot where the highest stored energy is recorded in LHD. 
Three
negative ion source NB are injected step by step and the repetitive pellet is injected 
to increase the density. The total injected NB power attained 9 MW in this shot. 
The line averaged density once reaches 2.0 $\times$ 10$^{20}$ m$^{-3}$ 
and decreases with the time constant of 0.5 s.
The 168 GHz power of the level of 800 kW in total is applied on such high density plasma.
dependence of heating efficiency of 168 GHz on target plasma density.
The heating efficiency is estimated from the difference of the time derivative of
stored energy just before and after the turning on and off time of 168 GHz pulse.
\Fref{fig:Efficiency} b)
shows such estimated heating efficiency as a function of
line-averaged density. Difference of the mark corresponds to the timing of turing on and off for
stored energy.  Appreciable heating effect is observed for the shots of the averaged electron density
above 1.0 $\times$ 10$^{20}$ m$^{-3}$.  Although the degradation of the efficiency appears above 
1.0 $\times$ 10$^{20}$ m$^{-3}$, heating efficiency of near 30\% is observed near the
right hand cutoff density of 1.75 $\times$ 10$^{20}$ m$^{-3}$. Highest efficiency point for
1.7 and 1.2 $\times$ 10$^{20}$ m$^{-3}$ just corresponds to the turn on and off data 
from Fig. \ref{fig:Efficiency} a).
This figure clearly shows the effectiveness of second harmonic heating.
%
    \section{ Long Pulse Operation}
 Demonstration of steady state high Te plasma sustainment is 
  important from the view point of extending the helical magnetic 
  confinement device to the fusion reactor. 
  Technical issue in the 
  CW ECH system is also important in the 
  next step fusion device as ITER.
  Especially, the heat removal around gyrotron, 
  MOU and transmission components are of critical issues. 
  Development of low loss 
  and safety component is required.
  Management of heat flux into the diverter and first wall,
   plasma density, temperature and particle control and sustainment 
  in the steady state condition are to be established. 
  A CW gyrotron with the diamond window at the frequency of 84 GHz is 
  introduced to perform the steady state plasma sustainment in LHD.
  This gyrotron is connected to one of the evacuated corrugated waveguide 
  systems of 31.75 mm id. (\#5) listed in Table \ref{tab:TransList}.

  In order to monitor the temperature rise in the waveguide components, 
  multi-point temperature measurement system is introduced and attached 
  the sensors along the long transmission line.
  Since existing focusing antenna for this line had no cooling capability,
  a corrugated up taper from 31.75 mm id to 63.5 mm id is connected
  and the output beam is directed to a magnetic axis.
  Because of limited injection power, gas feed rate is controlled 
  by repetitive pulse puff.
  Almost all gyrotron parameters stayed almost constant during 766 sec 
injection.
  Collector, body voltages are well controlled and kept constant 
during the shot. 
  Beam and body current and little increases with the time constant of 300 s 
  and then saturates to be constant toward the end of the pulse.
  Temperature difference between inlet and outlet of 
  several channels of 
  the gyrotron 
  coolant are still a little increasing even after such long pulse operation,
  but demonstrated that this gyrotron can be operated more than 1000 s with 
  the power level at 150 kW. 
  Operation of the interlock on the pressure in the waveguides due to the 
  excess outgas  resulted from the temperature 
  rise had been the major
  cause of the termination of  the pulse.
  Out gassing were far from saturation and the pressure kept 
  increased due to the 
  poor conductance in the waveguide.
  Figure \ref{fig:CWplasma} shows time evolutions of a) 
  injection power, b) averaged
  electron density, c) ECE radiation temperature, d) 
  radiation power and e) temperature
  rise in waveguide component. The repetitive gas puff 
  is controlled not to 
  cause radiation collapse. This is the reason why the density, 
  temperature and radiation
  fluctuates.  In average, plasma with electron density of 
  2.4 $\times $ 10$^{17}$ m$^{-3}$, 
  and ECE radiation temperature of 240 eV 
  ( this is not necessarily the real electron temperature
  due optically thin) is sustained by 72 kW injection power.  
  It is noted that the density decay rate just after 
  the gas puffing pulse decreases
  towards the end of the pulse. This might reflect 
  the change of recycling from the wall.
  It is necessary to continue the effort to increase the injection power by
  enforcing the cooling and pumping of the waveguide.
  
   \section {Conclusions}

  ECRH system have been upgraded step by step to 
  inject total power more than 2 MW.
  The optimizations of the injection condition for 
  almost all antennas are performed
  in the beam steering angle and also polarization.
  The power deposition profile is carefully controlled and utilized to 
  achieve high electron temperature of more than 10 keV 
  and formation of ITB. Sharp power deposition is 
  also utilized  in analyzing and understanding the transport mechanisms.
  Second harmonic heating of 168 GHz demonstrated effective heating
  at the density of more than 1.5 $\times$ 10$^{20}$ m$^{-3}$. 
  Long pulse more than 12 minutes of 70 kW at 84 GHz 
  is injected and sustained the 
  plasma for 756 sec at the averaged density of 2.4 $\times$ 10$^{17}$ m$^{-3}$. 
  
  \ack
  Technical stuffs of ECRH group and LHD group are greatly appreciated for their
  efforts to operate ECRH systems and LHD. 
  Authors would like to acknowledge cooperation of the GYCOM 
  to operate high power and CW gyrotron.

\end{document}